\documentclass[aps,prd,preprintnumbers,showpacs,showkeys,nofootinbib,
superscriptaddress,fleqn,floatfix,tightenlines]{revtex4-1}
\usepackage{amsmath,amsfonts,amssymb,amscd,amsxtra,amsthm}
\usepackage{graphicx}  % Include figure files
\usepackage{epstopdf}
\usepackage{dcolumn}  % Align table columns on decimal point
\usepackage{bm}          % bold math 
\usepackage{slashed}
\usepackage[utf8]{inputenc} 
\usepackage{booktabs}
\usepackage[normalem]{ulem} % \sout{old text} for strikeout
\usepackage[dvipsnames]{xcolor} % For blue in-text comments and
\usepackage{tabularx}
\usepackage{enumitem}  
\usepackage{array} 
\usepackage{slashed}
\usepackage{tikz}
\usepackage{cleveref} 
\usepackage{multirow}
\renewcommand\sout{\bgroup \color{red} \ULdepth=-.5ex \ULset}

\makeatletter

\begin{document}  
\preprint{INHA-NTG-01/2018}
\title{Mass spectra of singly heavy baryons \\ in a self-consistent
  chiral quark-soliton model}    
%--------------------------------------------------
\author{June-Young Kim}
\email[E-mail: ]{junyoung.kim@inha.edu}
\affiliation{Department of Physics, Inha University, Incheon 22212,
Republic of Korea}
%--------------------------------------------------
\author{Hyun-Chul Kim}
\email[E-mail: ]{hchkim@inha.ac.kr}
\affiliation{Department of Physics, Inha University, Incheon 22212,
Republic of Korea}
\affiliation{Research Center for Nuclear Physics (RCNP), Osaka
University, Ibaraki, Osaka 567-0047, Japan}
\affiliation{School of Physics, Korea Institute for Advanced Study 
  (KIAS), Seoul 02455, Republic of Korea}
%--------------------------------------------------
\author{Ghil-Seok Yang}
\email[E-mail: ]{ghsyang@ssu.ac.kr}
\affiliation{Department of Physics, Soongsil University, Seoul 06978,
Republic of Korea}
%--------------------------------------------------
\date{\today}
\begin{abstract}
We investigate the mass spectra of the lowest-lying singly heavy
baryons, based on the self-consistent chiral quark-soliton model. 
We take into account the rotational $1/N_c$ and strange current quark
mass ($m_{\mathrm{s}}$) corrections.  Regarding $m_{\mathrm{s}}$ as a
small perturbation, we expand the effective chiral action to the
second order with respect to $m_{s}$. The mass spectra of heavy
baryons are computed and compared with the experimental 
data. Fitting the classical masses of the heavy baryon to the center
mass of each representation, we determine the masses of all the
lowest-lying singly heavy baryons. We predict the mass of the
$\Omega_b^*$ baryon to be 6081.9 MeV, when the second-order
$m_{\mathrm{s}}$ corrections are included.    
\end{abstract}
\pacs{}
\keywords{Heavy baryons, pion mean fields, 
chiral quark-soliton model, flavor SU(3) symmetry breaking} 
\maketitle
%--------------------------------------------------
\section{Introduction}
\begin{figure}[htp]
\centering
\includegraphics[scale=0.8]{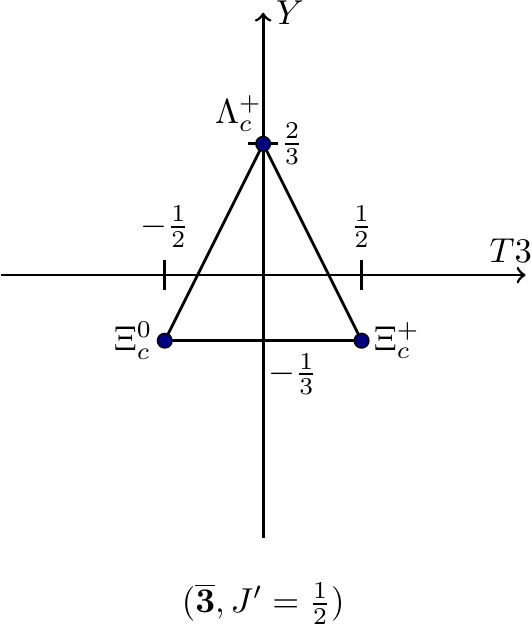}\qquad
\includegraphics[scale=0.8]{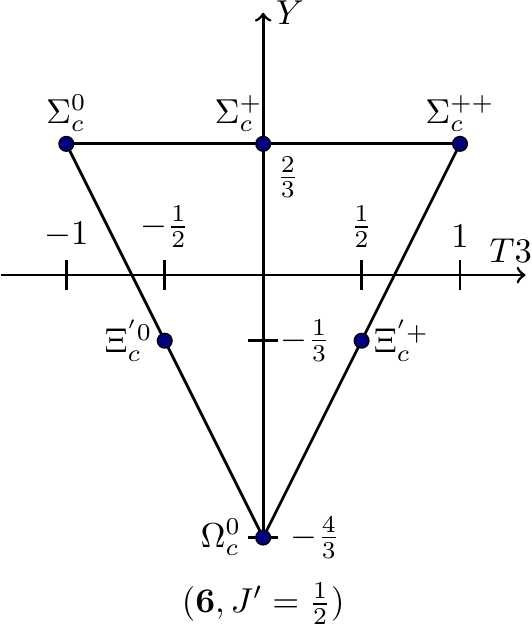}\qquad
\includegraphics[scale=0.8]{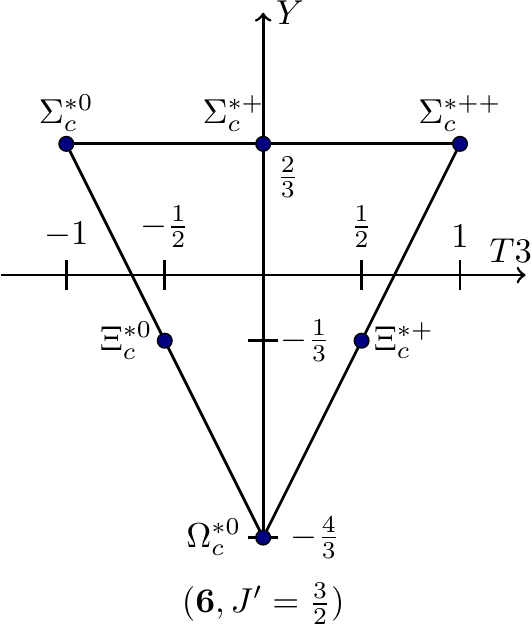}
\caption{The anti-triplet ($\overline{\bm{3}}$) and sextet ($\bm{6}$)
  representations of the lowest-lying heavy baryons. The left panel
  draws the weight diagram for the anti-triplet with the total spin
  $\frac{1}{2}$. The centered panel corresponds to that for the sextet
  with the total spin $1/2$ and the right panel depicts that
  for the sextet with the total spin $3/2$.} 
\label{fig:1}
\end{figure}

%--------------------------------------------------
Interest in heavy baryons is renewed as a series of new experimental 
data on them was 
reported~\cite{Aaltonen:2007ar, Chatrchyan:2012ni, Abazov:2008qm, 
  Kuhr:2011up, Aaij:2012da, Aaij:2013qja, Aaij:2014esa, Aaij:2014lxa,
  Aaij:2014yka, Aaij:2017nav}. A conventional heavy baryon is composed
of a heavy quark and two light quarks. Since the mass of the heavy
quark is very large in comparison with that of the light quarks, we
can take the limit of the infinitely heavy mass of the heavy 
quark, i.e. $m_Q\to \infty$. It leads to the conservation of the
heavy-quark spin $\bm{J}_Q$, which results also in the 
conservation of the total spin of light quarks: $\bm{J}
\equiv \bm{J}'-\bm{J}_Q$, where $\bm{J}'$ is the spin of the heavy
baryon~\cite{Isgur:1989vq, Isgur:1991wq, Georgi:1990um}. This is
called heavy-quark spin symmetry that makes $J$ a good
quantum number. In this limit, the heavy quark can be regarded merely
as a static color source and dynamics inside a heavy baryon is mostly 
governed by the light quarks consisting of it. Thus the two light
quarks determine which flavor $\mathrm{SU(3)}_{\mathrm{f}}$
representation a heavy baryon belongs to. There are two different
representations: $\bm{3}\otimes \bm{3}=\overline{\bm{3}} \oplus
\bm{6}$. The anti-triplet ($\overline{\bm{3}}$) has $J=0$
and total $J'=1/2$ whereas the sextet ($\bm{6}$) has
$J=1$. Thus, the spin of a heavy baryon is determined by
the spin alignment of the light-quark pair together with a heavy 
quark. It becomes either $J'=1/2$ or $J'=3/2$. So, there are 15
different lowest-lying heavy baryons classified as shown in
Fig.~\ref{fig:1} in the case of charmed baryons.  

The masses of singly heavy baryons have been studied within various
chiral solitonic models, in particular, based on bound-state
approaches~\cite{Callan:1985hy, Callan:1987xt}. The model was
originally applied to hyperons, the strange quark being regarded as a
heavy one.  This bound-state approach was employed to describe charmed
baryons as soliton-$D$ meson bound states~\cite{Rho:1990uy}. In the
advent of heavy quark symmetry, References~\cite{Jenkins:1992zx,
  Guralnik:1992dj, Gupta:1993kd, Momen:1993ax} 
incorporated this symmetry and describe singly heavy baryons as a
bound state of a soliton and heavy mesons. Moreover, in the original
bound-state approach, the whole soliton-heavy meson bound state was
quantized collectively, whereas Refs.~\cite{Jenkins:1992zx,
  Guralnik:1992dj, Gupta:1993kd, Momen:1993ax} first quantized the
soliton and then coupled it to heavy mesons. In Ref.~\cite{Oh:1994yv},
it was shown that these two different quantization schemes in the
bound-state approach are in fact equivalent. 

In the chiral quark-soliton model ($\chi$QSM), singly heavy baryons
were examined only very recently. Reference~\cite{Yang:2016qdz} put
forward a mean-field approach to describe the masses of singly heavy
baryons, being motivated by Ref.~\cite{Diakonov:2010tf}. 
The main idea of this mean-field approach is rooted in
 Refs.~\cite{Witten:1979kh,Witten:1983}, where Witten has 
 suggested that in the limit of the large number of colors ($N_c$) the 
 nucleon can be viewed as a bound state of $N_c$ 
 \textit{valence} quarks in a pion mean field with a hedgehog
 symmetry~\cite{Pauli:1942kwa, Skyrme:1961vq}, as the quantum
 fluctuation around the saddle point of the pion field is 
 $1/N_c$ suppressed. In this large $N_c$ limit, the presence of $N_c$ 
\textit{valence} quarks that constitute the lowest-lying baryons
brings about the vacuum polarization, which  
produces the pion mean field. The $N_c$ valence quarks are also 
\emph{self-consistently} influenced by this pion mean field. 
Because of the hedgehog symmetry, an $\mathrm{SU(2)}$ soliton is
embedded into the isospin subgroup of the flavor
$\mathrm{SU(3)}_{\mathrm{f}}$~\cite{Witten:1983}, which was also 
employed by various chiral soliton models~\cite{Guadagnini:1983uv,
  Mazur:1984yf, Jain:1984gp}. The collective quantization of the
chiral soliton yields the collective Hamiltonian with effects of
flavor $\mathrm{SU(3)}_{\mathrm{f}}$ symmetry breaking.  This
mean-field approach is called the $\chi$QSM~\cite{Diakonov:1987ty,
  Christov:1995vm, Diakonov:1997sj}. One salient feature of the
$\chi$QSM is that the right hypercharge is constrained to be
$Y'=N_c/3$ imposed by the $N_c$ valence quarks. This right hypercharge
selects allowed representations of light baryons such as the baryon
octet ($\bm{8}$), the decuplet ($\bm{10}$), etc. The $\chi$QSM
described successfully the properties of the lowest-lying light
baryons such as the mass splittings~\cite{Blotz:1992pw}, the form
factors~\cite{Kim:1995mr, Silva:2001st, Ledwig:2008ku, Ledwig:2010tu},
and parton distributions~\cite{Diakonov:1996sr}. 

In the present work, we investigate the mass spectra of singly 
heavy baryons in the ground states within the framework of the
$\chi$QSM. Since a singly heavy baryon contains $N_c-1$ light valence
quarks, the imposed constraint $Y'$ should be modified as
$\overline{Y} =(N_c-1)/3$. This allows the lowest-lying
representations: the baryon anti-triplet ($\bar{\bm{3}}$) and the
baryon sextet ($\bm{6}$). While in Ref.~\cite{Yang:2016qdz} all
dynamical parameters were fixed by using the experimental data, we
will compute them here explicitly in a self-consistent way. This
explicit calculation has a certain advantage over the previous
model-independent analysis. Since we calculate the valence and sea
contributions separately, we can correctly consider the pion mean
field that is produced only by the $N_c-1$ valence quarks whereas the
vacuum polarization is kept to be the same as in the case of light
baryons. On the other hand, the model calculation suffers from a
deficiency: the classical soliton mass turns out to be rather large in
the model, which is a usual problem in any chiral soliton models. It
means that the predicted values of baryon masses from the model tend
systematically to be overestimated. Thus, we will first concentrate on
the mass splittings of the lowest-lying heavy baryons in the present
work, focusing on the effects of SU(3) symmetry breaking. 

Regarding the mass of the strange current quark as a small
perturbation, we first consider its linear-order corrections to the
masses of heavy baryons and then take into account the second-order
corrections. However, a caveat on the second-order corrections should
be mentioned. In principle the effective chiral action may include a
term that is proportional to the square of the current quark
masses. However, so far any rigorous theoretical method for that is
not known. Thus, the second-order corrections in the present work
imply only the contributions arising from the second-order
perturbation theory. Bearing this warning in mind, we will examine the
masses of both the singly charmed and bottom baryons.
Taking a practical point of view, that is, we fix the center
masses in each representation by using the experimental data as in 
Ref.~\cite{Yang:2016qdz}. Then we are able to produce all the values
of the lowest-lying singly heavy baryons. We also predict the mass of
the $\Omega_b^*$ baryon, of which the value is experimentally yet
unknown. 

The structure of the present work is sketched as follows: In Section
II, we briefly review the $\chi$QSM for singly heavy baryons. In
Section III, we examine numerically the effects of
$\mathrm{SU(3)}_{\mathrm{f}}$ symmetry breaking. We then present the
prediction of the heavy baryon masses, fixing the center masses in
each representation by the data. The last Section is devoted to the
summary and conclusions of the present work. In Appendices, we have
compiled all necessary formulae explicitly.

%--------------------------------------------------
\section{General formalism\label{sec:2}}
%--------------------------------------------------
A heavy quark inside a heavy baryon can be regarded as a static color
source in the limit of the infinite heavy-quark mass $m_Q\to
\infty$. In this case, the heavy quark is only required to make the
heavy baryon a color singlet. So, it can be described as
the correlation function of the $N_c-1$ light-quark field operators in
Euclidean space, defined by  
\begin{align}
\Pi_{B}(0, T) = \langle J_B (0, T/2) J_B^\dagger
  (0,-T/2) \rangle_0 = \frac{1}{Z}\int  \mathcal{D} U
  \mathcal{D}\psi^\dagger 
  \mathcal{D}\psi J_B(0,T/2) J_B^\dagger (0,-T/2)
  e^{\int d^4 x\,\psi^\dagger (i\rlap{/}{\partial} + i
  MU^{\gamma_5}+ i \hat{m})\psi}  ,   
\label{eq:corr1}
\end{align}
where $J_B$ denotes the light-quark current consisting of $N_c-1$ 
light quarks for a heavy baryon $B$
\begin{align}
J_B(\bm{x}, t) = \frac1{(N_c-1)!}
  \varepsilon^{\beta_1\cdots\beta_{N_c-1}} \Gamma_{J'J_3',TT_3}^{\{f\}}
  \Psi_{\beta_1f_1}(\bm{x}, t) \cdots \Psi_{\beta_{N_c-1}f_{N_c-1}}
  (\bm{x}, t).      
\end{align}
$\beta_i$ represent color indices and 
$\Gamma_{J'J_3',TT_3}^{\{f_1\cdots f_{N_c-1}\}}$ is a matrix with both
flavor and spin indices. $J'$ and $T$ are the spin and isospin of the
heavy baryon, respectively. $J_3'$ and $T_3$ are their third
components, respectively. The notation $\langle \cdots \rangle_0$ in
Eq.~(\ref{eq:corr1}) stands for the vacuum expectation value. $M$
denotes the dynamical quark mass and $U^{\gamma_5}$ is defined as   
\begin{align}
U^{\gamma_5} = U\frac{1+\gamma_5}{2} + U^\dagger \frac{1-\gamma_5}{2},
\end{align}
with 
\begin{align}
U = \exp\left[i\frac{\pi^a \lambda^a}{f_\pi}\right].  
\end{align}
$\pi^a$ represents the pseudo-Goldstone field and $f_\pi$ is the pion
decay constannt. $\hat{m}$ is the flavor
matrix of the current quark masses, written as
$\hat{m}=\mathrm{diag}(m_{\mathrm{u}},\,m_{\mathrm{d}},\,m_{\mathrm{s}})$. We
assume in the present work isospin symmetry,
i.e. $m_{\mathrm{u}}=m_{\mathrm{d}}$. The strange current quark mass
will be treated perturbatively. 

Integrating over the quark fields, we obtain the correlation function
as 
\begin{align}
\Pi_{B}(0, T) =
  \frac{1}{Z}\Gamma_{J'J_3',TT_3}^{\{f\}}\Gamma_{J'J_3',TT_3}^{\{g\}*} \int
  \mathcal{D} U \prod_{i=1}^{N_c-1}  \left\langle 0,T/2\left|
  \frac1{D(U)} \right|0,-T/2\right\rangle
  e^{-S_{\mathrm{eff}}(U)}, 
\label{eq:corr2}
\end{align}  
where $D(U)$ is defined as
\begin{align}
D(U) = i\gamma_4 \partial_4 + i\gamma_k \partial_k + i MU^{\gamma_5} +
  i \hat{m},
\end{align}
and $S_{\mathrm{eff}}$ represents the effective chiral action written
as 
\begin{align}
S_{\mathrm{eff}} = -N_c \mathrm{Tr}\log D(U).  
\label{eq:effecXac}
\end{align}

The correlation function at large separation of the Euclidean time
$\tau$ picks up the ground-state energies~\cite{Diakonov:1987ty,
  Christov:1995vm} 
\begin{align}
\lim_{\tau\to\infty} \Pi_B(\tau) \sim \exp[-(N_c-1) E_{\mathrm{val}} +
  E_{\mathrm{sea}} \tau],   
\end{align}
where $E_{\mathrm{val}}$ and $E_{\mathrm{sea}}$ the valence and
sea quark energies. The soliton mass is then derived
by minimizing self-consistently the energies around the saddle point
of the chiral field $U$ 
\begin{align}\label{eq:sol}
\left.\frac{\delta}{\delta U}[ (N_c-1) E_{\mathrm{val}} +
  E_{\mathrm{sea}}]\right|_{U_c} = 0,    
\end{align}
which yields the soliton mass 
\begin{align}\label{eq:solnc}
M_{\mathrm{sol}} = (N_c-1) E_{\mathrm{val}}(U_c) + E_{\mathrm{sea}}(U_c).  
\end{align}
Since a singly heavy baryon contains the heavy quark, its classical
mass of a heavy baryon should be expressed as 
\begin{align}
M_{\mathrm{cl}} = M_{\mathrm{sol}} + m_Q,  
\label{eq:classical_mass}
\end{align}
where $m_Q$ is the \textit{effective} heavy quark mass that is
different from that discussed in QCD and will be absorbed in the
center mass of each representation, which will be discussed later.

As in the light-baryon sector, we expect that the lowest-lying heavy
baryons will arise from rotational excitations of the light-quark
soliton whereas the heavy quark is kept to be static. Keeping in mind
that the SU(2) soliton $U_c(\bm{r})$ has hedgehog symmetry, we embed
it into SU(3)~\cite{Witten:1983} 
\begin{align}
U(\bm{r}) = \begin{pmatrix}
U_c (\bm{r}) & 0 \\
0 & 1
\end{pmatrix}.
\label{eq:embed}
\end{align}
To find the $1/N_c$ quantum fluctuations, we need to integrate 
the meson fields over small oscillations of the $U(\bm{r})$ field
around the saddle point. However, we will not carry out this procedure
and this is often called the mean-field approximation. On the other
hand, we have to consider explicitly the rotational zero modes that
are not small and cannot be neglected. Thus, we restrict ourselves to
take into account these zero modes only.  Considering a slowly
rotating hedgehog field $U(\bm{r})$ in Eq.(\ref{eq:embed}) 
\begin{align}
U(\bm{r},\,t) = A(t) U(\bm{r}) A^\dagger (t),  
\end{align}
where $A(t)$ is an element of flavor SU(3) matrix, we can find the
collective Hamiltonian to describe heavy baryons. For a detailed
formalism of the semiclassical quantization, we refer to
Ref.~\cite{Christov:1995vm}. Regarding the angular velocity of the
soliton and the current strange quark mass as small parameters, we can
expand the quark propagator in Eq.(\ref{eq:corr2}) with respect to them.  

Having quantized the chiral soliton, we arrive at the collective
Hamiltonian for singly heavy baryons
\begin{align}
H =& H_{\mathrm{sym}} + H^{(1)}_{\mathrm{sb}} +  H^{(2)}_{\mathrm{sb}},
\end{align}
where $H_{\mathrm{sym}}$ represents the flavor SU(3) symmetric part, 
$H^{(1)}_{\mathrm{sb}}$ and $H^{(2)}_{\mathrm{sb}}$ the
SU(3) symmetry-breaking parts respectively to the first and second
orders, which will be discussed later. $H_{\mathrm{sym}}$ is expressed
as  
\begin{align} 
H_{\mathrm{sym}}=M_{\mathrm{cl}}+\frac{1}{2I_{1}}\sum_{i=1}^{3}
\hat{J}^{2}_{i} +\frac{1}{2I_{2}}\sum_{a=4}^{7}\hat{J}^{2}_{a},  
\end{align} 
where $I_{1}$ and $I_{2}$ denote the moments of inertia of the
soliton. The explicit expressions for $I_{1,\,2}$ are given in
Eq.~\eqref{eq:momIi}. The operators  $\hat{J}_{i}$ are the SU(3)
generators. In the $(p,\,q)$ representation of the SU(3) group, we
find the eigenvalue of the Casimir operator $\sum_{i=1}^8
J_i^2$ as    
\begin{align}
C_2(p,\,q) = \frac13 \left[p^2 +q^2 + pq + 3(p+q)\right].   
\end{align}
Thus, the eigenvalues of $H_{\mathrm{sym}}$ is obtained as  
\begin{align} 
E_{\mathrm{sym}}(p,q) = M_{\mathrm{cl}}+ \frac{1}{2I_{1}} J(J+1) 
+\frac{1}{2I_{2}}\left[C_2(p,\,q) - J(J+1)\right] 
-\frac{3}{8I_{2}}\overline{Y}^{2}.
\label{eq:RotEn}
\end{align} 
The right hypercharge $\overline{Y}$ is constrained to be $(N_c-1)/3$,
which is imposed by the $N_c-1$ valence quarks inside a singly heavy
baryon. The wavefunctions of the
singly heavy baryon are derived as 
\begin{align} 
\psi_B^{({\mathcal{R}})}(J'J'_3,J;A)=
\sum_{m_{3}=\pm1/2}C^{J' J'_3}_{J_{Q} m_{3} J 
J_{3}} \chi_{m_{3}} \sqrt{\mathrm{dim}(p,\,q)}
(-1)^{-\frac{ \overline{Y}  }{2}+J_3}
  D^{(\mathcal{R})\ast}_{(Y,T,T_3)(\overline{Y}  ,J,-J_3)}(A),  
\label{eq:waveftn}\end{align} 
where 
\begin{align}
\mathrm{dim}(p,\,q) = (p+1)(q+1)\left(1+\frac{p+q}{2}\right).  
\end{align}
Note that a similar expression can be found in
Ref.~\cite{Momen:1993ax}, though its formalism is rather different
from the present one. $J$ and $J_{Q}$ stand for the soliton spin and
heavy-quark spin, respectively. ${J_{3}}$ and ${m_{3}}$ represent the
corresponding third components, respectively. Since the spin operator
for the heavy baryon is given by the addition of the soliton and
heavy-quark spin operators 
\begin{align}
\label{eq:quantum}
\bm{J}'=\bm{J}_{Q}+\bm{J},
\end{align} 
the relevant Clebsch-Gordan coefficients appear in
Eq.(\ref{eq:waveftn}). The SU(3) Wigner $D$ function in
Eq.(\ref{eq:waveftn}) is just the wavefunction for the quantized
soliton consisting of the $N_c-1$ valence quarks, whereas $\chi_{m_3}$
is the Pauli spinor for the heavy quark. $\mathcal{R}$ stands for a
SU(3) irreducible representation corresponding to $(p,\,q)$. 

Since a singly heavy baryon consists of $N_c-1$ valence quarks, we
have two irreducible representations when $N_c=3$:
$\bm{3}\otimes\bm{3}=\overline{\bm{3}} \oplus \bm{6}$. Thus, we have
the following representations for the lowest-lying singly heavy
baryons 
\begin{align}
\left[ \ \overline{\bm{3}}_{0} \ \right ] &=D(0, 1)  
\mbox{ :the anti-triplet with $J=0$}, 
                    \cr   
\left[ \ \bm{6}_{1}  \ \right ] &=D(2, 0)  \mbox{ :the sextet with $J=1$}.
\end{align}
The soliton being coupled to the heavy quark, we finally get three
different representations, which have been illustrated already in
Fig.~\ref{fig:1}.  Since the soliton in the sextet ($J=1$) is coupled
to the heavy quark ($J_Q=1/2$), we have two sextet representations
with spin 1/2 and 3/2 respectively, which are degenerate. The 
hyperfine spin-spin interaction will lift this degeneracy.  

Since a singly heavy baryon consists of $N_c-1$ valence quarks, the
pion mean fields should be changed. In Refs.~\cite{Yang:2016qdz,
  Kim:2017jpx}, a scale factor was introduced to explain the
modification of the mean field, of which the value was taken to be in
the range of $1-0.66$. Because all dynamical variables being
proportional to the color factor were fixed by the experimental data
in Refs.~\cite{Yang:2016qdz, Kim:2017jpx}, it was impossible to
decompose the valence and sea parts. On the other hand, we can treat
separately the valence and sea quark contributions in the present
work. So, we will replace $N_c$ factor with $N_c-1$ only in front of the
valence part of the dynamical parameters, while we keep the sea part
intact. 

In order to describe the mass splittings of SU(3) baryons in a
specific representation, we have to consider the effects of flavor
SU(3) symmetry breaking, dealing with the mass of the strange current
quark, $m_{\mathrm{s}}$, as a small perturbation. First, we consider the
first-order corrections that are proportional to the linear $m_{\mathrm{s}}$, and 
then we proceed to take into account the second-order corrections.
In this case, the baryon wavefunctions are no more in pure states but
are mixed with those of higher representations.  Thus, there are
two different contributions: one from the collective Hamiltonian and
the other from the baryon wavefunctions. Both corrections will be
considered as second-order contributions.  

\subsection{Mass splittings  to the linear order}
The symmetry-breaking part of the collective
Hamiltonian is given as~\cite{Blotz:1992pw, Christov:1995vm}
\begin{align} 
H^{(1)}_{\mathrm{sb}} 
=&\frac{\overline{\Sigma}_{\pi N}}{m_{0}}\frac{m_{\mathrm{s}}}{3}
+\alpha D^{(8)}_{88}+ \beta \hat{Y}
+ \frac{\gamma}{\sqrt{3}}\sum_{i=1}^{3}D^{(8)}_{8i}
\hat{J}_{i},
\label{sb}
\end{align}
where
\begin{align} 
\alpha=\left (-\frac{\overline{\Sigma}_{\pi N}}{3m_0}+\frac{
  K_{2}}{I_{2}}\overline{Y}  
\right )m_{\mathrm{s}},
 \;\;\;  \beta=-\frac{ K_{2}}{I_{2}}m_{\mathrm{s}}, 
\;\;\;  \gamma=2\left ( \frac{K_{1}}{I_{1}}-\frac{K_{2}}{I_{2}} 
 \right ) m_{\mathrm{s}}.
\label{eq:alphaetc}
\end{align}
The first term in Eq.~\eqref{sb} can be absorbed into the symmetric
 part of the Hamiltonian, since it does not contribute to the mass
 splittings of heavy baryons in a given representation. 
The $m_0$ represents the averaged mass of the up and down
quarks. The three parameters $\alpha$, $\beta$, and $\gamma$ are
expressed in terms of the moments of inertia $I_{1,\,2}$ and
$K_{1,\,2}$, of which the valence parts are  different from those in
the light baryon sector by the color factor $N_c-1$ in place of
$N_c$. The valence part of $\overline{\Sigma}_{\pi N}$ is different
from the $\pi N$ sigma term by the prefactor $N_c-1$, that is,
$\overline{\Sigma}_{\pi N} = (N_c-1)N_c^{-1} \Sigma_{\pi 
   N}$, where $\Sigma_{\pi N} = (m_u+m_d)\langle N| \bar{u} u +
 \bar{d} d|N\rangle = (m_u+m_d)  \sigma$. The explicit expressions for
 the moments of inertia and the $\pi N$ sigma term can be found in
 Appendix~\ref{app:A}. Note that their sea parts are the same as in the light
baryon sector. 

Taking into account the $m_{\mathrm{s}}$ corrections to the first order, we can
write the masses of the singly heavy baryons in representation
$\mathcal{R}$ as  
\begin{align}
M_{B,\mathcal{R}}^Q = M_{\mathcal{R}}^Q + M_{B,\mathcal{R}}^{(1)},  
\label{eq:FirstOrderMass}
\end{align}
where 
\begin{align}
M_{\mathcal{R}}^Q = m_Q + E_{\mathrm{sym}}(p,q).  
\label{eq:center_mass}
\end{align}
$M_{\mathcal{R}}^Q$ is called the center mass of a heavy baryon in
representation $\mathcal{R}$. $E_{\mathrm{sym}}(p,q)$ is defined in
Eq.~\eqref{eq:RotEn}.  
Note that the lower index $B$ denotes a certain baryon belonging to a
specific representation $\mathcal{R}$. The upper index $Q$ stands for
either the charm sector ($Q=c$) or the bottom sector ($Q=b$).
The center masses for the anti-triplet and sextet representations can
be explicitly written as
\begin{align}
\label{eq:center_mass}
M_{\overline{\bm{3}}}^Q = M_{\mathrm{cl}} + \frac1{2I_2}, \;\;\; 
M_{\bm{6}}^Q = M_{\overline{\bm{3}}}^Q +  \frac1{I_1},
\end{align}
where $M_{\mathrm{cl}}$ was defined in Eq.~\eqref{eq:classical_mass}.
The second term in Eq.~(\ref{eq:FirstOrderMass}) denotes the
linear-order $m_{\mathrm{s}}$ corrections to the heavy baryon mass 
\begin{align}
M^{(1)}_{B,{\cal{R}}} = \langle B, {\cal{R}} | H_{\mathrm{sb}}^{(1)} 
| B, {\cal{R}} \rangle  = Y\delta_{{\cal{R}}},
\end{align}
where
 \begin{align} 
&\delta_{\overline{\bm{3}}}=\frac{3}{8}\alpha+\beta, \;\;\;\;
\delta_{\bm{6}}=\frac{3}{20}\alpha+\beta-\frac{3}{10}\gamma. 
\end{align}
The values of the matrix elements for the relevant SU(3) Wigner $D$
functions are tabulated in Appendix~\ref{app:b}.
Thus, we obtain the masses of the lowest-lying singly heavy baryons
\begin{align}
M_{B,\overline{\bm{3}}}^Q = M_{\overline{\bm{3}}}^Q  +
                     Y \delta_{\overline{\bm{3}}} ,\;\;\;
M_{B,\bm{6}}^Q =M_{\bm{6}}^Q  + Y  \delta_{\bm{6}}, 
  \label{eq:firstms}
\end{align}
with the linear-order $m_{\mathrm{s}}$ corrections taken into account. 
\subsection{Mass splittings  to the second order}
We now consider the second-order $m_{\mathrm{s}}$ corrections. When we include
the second-order corrections, the collective wavefunction of baryons
is no more in a pure state but is mixed with those in 
higher representations. Using the standard method of perturbation
theory, we can derive the second-order $m_{\mathrm{s}}$ corrections to the baryon
mass, which arise from the baryon wavefunctions~\cite{Blotz:1993cc} 
\begin{align}
M^{(2)\mathrm{(wf)}}_B=\sum_{\cal{R} \neq \cal{R'}}\frac{\left | \langle
 {\cal{R'}},B|H^{(1)}_{\mathrm{sb}}|{\cal{R}},B\rangle \right |^2}
 {M_{\mathcal{R}}^Q-M_{\mathcal{R'}}^Q},
\end{align}
where ${\mathcal{R}}'$ denote higher representations that are
different from $\mathcal{R}$. These representations are determined by
the irreducible decomposition of the following products
$\overline{\bm{3}} \otimes \bm{8} = \overline{\bm{3}} \oplus \bm{6}
\oplus \overline{\bm{15}}$  and $\bm{6} \otimes \bm{8} =
\overline{\bm{3}} \oplus \bm{6} \oplus \overline{\bm{15}} \oplus
{\overline{\bm{24}}}$. The corresponding baryon wavefunction is then
expressed as a mixed state with those in higher representations 
\begin{align}
|B^{({\cal{R}})} \rangle=|{\cal{R}},B\rangle-\sum_{\cal{R} \neq \cal{R'}
}\frac{ \left |\langle {\cal{R'}},B|H^{(1)}_{\mathrm{sb}}|{\cal{R}},B
\rangle \right|}{M_{\mathcal{R}}^Q- M_{\mathcal{R'}}^Q}
|{\cal{R'}},B\rangle.  
\end{align}
Explicit calculation yields the collective wavefunctions of the
baryon anti-triplet and sextet, respectively, as  
\begin{align}
&|B_{\overline{\bm3}_{0}}\rangle = |\overline{\bm3}_{0},B\rangle + 
p^{B}_{\overline{15}}|\overline{\bm{15}}_{0},B\rangle, \cr
&|B_{\bm6_{1}}\rangle = |{\bm6}_{1},B\rangle +
  q^{B}_{\overline{15}}|{\overline{\bm{15}}}_{1},B 
\rangle + q^{B}_{\overline{24}}|{
{\overline{\bm{24}}}_{1}},B\rangle,
\end{align}
with the mixing coefficients
\begin{align}
&p^{B}_{\overline{15}}=p_{\overline{15}}\left[ \begin{array}{c}
2 \\
\sqrt{3}
\end{array} \right],\hspace{0.6cm}
q^{B}_{\overline{15}}=q_{\overline{15}}\left[ \begin{array}{c}
2\sqrt{2} \\
\sqrt{3} \\
0 
\end{array} \right],\hspace{0.6cm}
q^{B}_{\overline{24}}=q_{\overline{24}}
\left[ \begin{array}{c}
1 \\
\sqrt{3/2} \\
\sqrt{3/2} 
\end{array} \right],
\end{align}
where
\begin{align}
p_{\overline{15}}=-\frac{3}{16\sqrt{5}}\alpha I_{2},\hspace{0.5cm}
q_{\overline{15}}=-\frac{1}{4\sqrt{5}}(\alpha+\frac{2}{3}\gamma) I_{2},
\hspace{0.5cm} q_{\overline{24}}=-\frac{2}{25}(\alpha-\frac{1}{3}\gamma)I_{2}
,\hspace{0.2cm} 
\end{align}
in the bases of $[\Lambda_{Q},\Xi_{Q}]$ and
$[\Sigma_{Q},\Xi'_{Q},\Omega_{Q}]$, respectively. Then, we obtain the  
second-order corrections to the masses of the singly heavy baryons
from the baryon wavefunctions as 
\begin{align}
M^{(2){\mathrm{(wf)}}}_{\Lambda_Q}&=-I_2\frac{9}{160}\alpha^{2}, \cr
M^{(2){\mathrm{(wf)}}}_{\Xi_Q}&=-I_2\frac{27}{640}\alpha^{2},     \cr
M^{(2){\mathrm{(wf)}}}_{\Sigma_Q}&=-I_2\frac{1}{90}
(3\alpha+2\gamma)^{2}-I_2 \frac{2}{1125}(3\alpha-\gamma)^{2},\cr
M^{(2){\mathrm{(wf)}}}_{\Xi'_Q}&=-I_2\frac{1}{240}
(3\alpha+2\gamma)^{2}-I_2 \frac{1}{375}(3\alpha-\gamma)^{2}, \cr
M^{(2){\mathrm{(wf)}}}_{\Omega_Q}&=-I_2\frac{1}{375}
(3\alpha-\gamma)^{2} .
\end{align}

There are yet another second-order $m_{\mathrm{s}}$ corrections that come from the
collective Hamiltonian~\cite{Blotz:1993cc, Christov:1995vm}: 
\begin{align}
H^{(2)}_{\mathrm{sb}} = & m_{\mathrm{s}}^{2} \bigg{[} \frac{2}{3} 
\frac{K^{2}_{1}}{I_{1}} \sum^{3}_{i=1} D^{(8)}_{8i} (A) D^{(8)}_{8i} 
(A) +  \frac{2}{3} \frac{K^{2}_{2}}{I_{2}}\sum^{7}_{a=4}D^{(8)}_{8a}
(A)D^{(8)}_{8a}(A) \cr  &-  \frac{2}{3} N_{1} \sum^{3}_{i=1} 
D^{(8)}_{8i} (A) D^{(8)}_{8i} (A) -  \frac{2}{3} N_{2}\sum^{7}_{a=4}
D^{(8)}_{8a}(A)D^{(8)}_{8a}(A) -  \frac{2}{9} N_{0} \bigg{(}1-
D^{(8)}_{88}(A)\bigg{)}^2 \bigg{]},
\label{eq:2ndmsH}
\end{align}
where $N_0$, $N_1$, and $N_2$ are defined in Appendix~\ref{app:A}. 
Computing the matrix elements of Eq.~(\ref{eq:2ndmsH}), we obtain the
second-order $m_{\mathrm{s}}$ corrections to the masses of the singly heavy
baryons, which arise from the collective Hamiltonian   
\begin{align}
M^{(2){\mathrm{(op)}}}_{\Lambda_Q}& = m_{\mathrm{s}}^{2} \left (  \frac{3}{20}
\frac{K^{2}_{1}}{I_{1}}  +  \frac{2}{5} \frac{K^{2}_{2}}{I_{2}} 
+  \frac{13}{180} N_{0}  - \frac{3}{20}  N_{1} -  \frac{2}{5} N_{2} 
\right )  , \cr
M^{(2){\mathrm{(op)}}}_{\Xi_Q}& = m_{\mathrm{s}}^{2} \left (  \frac{3}{10}
\frac{K^{2}_{1}}{I_{1}}  +  \frac{3}{10} \frac{K^{2}_{2}}{I_{2}} 
-  \frac{7}{90} N_{0}  - \frac{3}{10}  N_{1} -  \frac{3}{10} N_{2} 
\right )  , \cr
M^{(2){\mathrm{(op)}}}_{\Sigma_Q}& = m_{\mathrm{s}}^{2} \left (  \frac{19}{90}
\frac{K^{2}_{1}}{I_{1}}  +  \frac{16}{45} \frac{K^{2}_{2}}{I_{2}} 
+  \frac{1}{90} N_{0}  - \frac{19}{90}  N_{1} -  \frac{16}{45} 
N_{2} \right )  , \cr
M^{(2){\mathrm{(op)}}}_{\Xi'_Q}& = m_{\mathrm{s}}^{2} \left (  \frac{4}{15}
\frac{K^{2}_{1}}{I_{1}}  +  \frac{1}{3} \frac{K^{2}_{2}}{I_{2}} 
-  \frac{2}{45} N_{0}  - \frac{4}{15}  N_{1} -  \frac{1}{3} N_{2} 
\right )  , \cr
M^{(2){\mathrm{(op)}}}_{\Omega_Q}& = m_{\mathrm{s}}^{2} \left (  \frac{1}{3}
\frac{K^{2}_{1}}{I_{1}}  +  \frac{4}{15} \frac{K^{2}_{2}}{I_{2}} 
-  \frac{1}{9} N_{0}  - \frac{1}{3}  N_{1} -  \frac{4}{15} N_{2} 
\right )\,  .
\end{align}
We will call them as the second-order $m_{\mathrm{s}}$ corrections
from the operator, so that we distinguish them from those coming from
the wavefunction corrections. Considering these second-order $m_{\mathrm{s}}$
corrections, we can extend Eq.~(\ref{eq:FirstOrderMass}) to 
\begin{align}
M_{B,\mathcal{R}}^Q = M_{\mathcal{R}}^Q + M_{B,\mathcal{R}}^{(1)} +
  M_{B,\mathcal{R}}^{(2)},    
\label{eq:full_mass}
\end{align}
where $M_{B,\mathcal{R}}^{(2)}$ denote the second-order corrections to 
a baryon in representation $\mathcal{R}$. 

%==========================================
\section{Results and discussion}
%==========================================
We are now in a position to compute the mass splittings of the
lowest-lying singly heavy baryons. Reference~\cite{Blotz:1992pw}
showed in detail how model parameters such as the cutoff masses and
the current quark masses can be fixed in the vacuum sector. In the
present work, we choose the constituent quark mass $M=420$ MeV,  
which provided the best prediction of baryon
observables~\cite{Christov:1995vm}. The mass of the strange current
quark, $m_s$, was taken to be 180 MeV also in previous works, since it
describes the mass splittings of the baryon octet and decuplet. In
fact, the value of the $m_s$ can be fixed by fitting the mass
splittings of the singly heavy baryon antitriplet and
sextet. The smaller values of $m_s$ yield the better results of the
mass splittings of the singly heavy baryons in comparison with those
of the baryon octet and decuplet. In the bottom baryon sector, even
the smaller value of $m_s$ is favored. Though it is an interesting
theoretical issue of understanding the reason for the dropping of the
$m_s$ value in the heavy baryon sector, we will use the canonical
value of $m_s=180$ MeV as in the previous
works~\cite{Kim:1995mr, Christov:1995vm, Silva:2001st, Ledwig:2008ku}.    

We follow Refs.~\cite{Blotz:1992pw, Christov:1995vm} for the numerical
methods of diagonalizing the Dirac equation in the presence of the
pion field and deriving the self-consistent solutions of the equations
of motion. However, we use a much larger size of the box in solving the
one-body Dirac equation such that we are able to reduce a numerical
instability and uncertainties~\footnote{10 fm is taken for the box
  size in the present work whereas 5 fm was used in
  Ref.~\cite{Blotz:1992pw}.}. Detailed numerical techniques and
relevant references are also given in Ref. ~\cite{Blotz:1992pw,
  Christov:1995vm}.   

\begin{table}[htp]
\caption{Numerical results of the moments of inertia, the $\Sigma_{\pi
    N}$ term, and the classical mass of the soliton. Note that the
  valence part of the moments of inertia for singly heavy baryons
  have the $N_c-1$ factor, whereas $N_c$ for light baryons.}  
\label{tab:1}
\centering 
\begin{tabular}{ c  c  | c c }
\hline
\hline 
Light baryon &  & Singly heavy
baryon &    \\ 
\hline
${I}_1$[fm] &1.108& $I_1$[fm] & {0.844}   \\ 
${I}_2$[fm] &0.529& $I_2$[fm] & {0.404} \\ 
${K}_1$[fm] &0.428 & $K_1$[fm] & {0.286}  \\ 
${K}_2$[fm] &0.272  & $K_2$[fm] & {0.181}  \\
${N}_0$[fm] &0.457 & $N_0$[fm] & {0.499}  \\
${N}_1$[fm] &0.410  & $N_1$[fm] & {0.380}  \\
${N}_2$[fm] &0.323  & $N_2$[fm] & {0.286}  \\
${\Sigma}_{\pi N}$[MeV] & 43.7 & 
$\overline{\Sigma}_{\pi N}$[MeV]& {40.0}  \\ 
${M}_{\mathrm{sol}}$[MeV] &1291.8  & $M_{\mathrm{sol}}$[MeV] &  {1093.3}  \\ 
\hline
\hline
\end{tabular}
\end{table}
In Table~\ref{tab:1} we list the numerical results of the moments of
inertia, the $\pi N$ sigma term, and the classical soliton mass 
$M_{\mathrm{sol}}$. As discussed in Section~\ref{sec:2}, the
expressions for the valence parts of all relevant quantities should be 
modified. The prefactor $N_c$ in those expressions for light
baryons, which counts the number of valence quarks, should be replaced
by the factor $N_c-1$, since a singly heavy baryon consists of $N_c-1$
light valence quarks. So, the difference between the left panel of
Table~\ref{tab:1} and the right one arises from the different
prefactor of  each valence part. The definition of
$\overline{\Sigma}_{\pi N}$ is just the same as $\Sigma_{\pi N}$
except for the valence contribution as shown in
Eq.~(\ref{eq:app1}). 

Though we are not able to determine the masses of singly heavy
baryons, because the center mass given in Eq.~\eqref{eq:center_mass}
in each representation seems overestimated, compared with the
experimental data. In addition, we must know the hyperfine interaction
which will lift the degeneracy of different spin states in the sextet 
representation. Thus, we will fix each center mass and parameters for
the hyperfine splitting, using the experimental data such that we can
determine the masses of the lowest-lying singly heavy baryons. We will
follow the method proposed by Ref.~\cite{Yang:2016qdz} in which the
spin-spin interaction Hamiltonian is given as 
\begin{align}
H_{solQ} = \frac23\frac{\kappa}{m_Q M_{\mathrm{sol}}} \bm{J}\cdot
   \bm{J}_Q = \frac23 \frac{\varkappa}{m_Q} \bm{J}\cdot
   \bm{J}_Q,
 \label{eq:hyperf_H}
\end{align}
where $\kappa$ represents the flavor-independent hyperfine coupling
constant. Note that the baryon anti-triplet does not acquire any
contribution from the hyperfine interaction, since the corresponding
soliton has spin $J=0$. On the other hand, the baryon sextet has $J=1$. Being
coupled to the heavy quark spin, it produces two different multiplets,
i.e., $J'=1/2$ and $J'=3/2$, of which the masses are expressed
respectively as  
\begin{align}
  \label{eq:hyperf_Masses}
 M_{B,\bm{6}_{1/2}}^Q = M_{B,\bm{6}}^Q -\frac23
  \frac{\varkappa}{m_Q},\;\;\;  
M_{B,\bm{6}_{3/2}}^Q =   M_{B,\bm{6}}^Q +\frac13
  \frac{\varkappa}{m_Q}.
\end{align}
Thus, we find the hyperfine mass splitting as  
\begin{align} 
M_{B,\bm{6}_{3/2}}^Q - M_{B,\bm{6}_{1/2}}^Q  =
  \frac{\varkappa}{m_Q}, 
\end{align}
where the corresponding numerical value can be determined by using the
center value of the sextet masses. In the charmed and bottom baryon
sectors, we obtain the corresponding numerical values respectively
\begin{align}
\frac{\varkappa}{m_c} =  68.1\,  \mathrm{MeV}, \;\;\;
\frac{\varkappa}{m_b} = 20.3\,  \mathrm{MeV}.  
\end{align}
Combining Eq.~\eqref{eq:hyperf_Masses} with Eq.~\eqref{eq:full_mass},
we can derive the final masses of the lowest-lying singly heavy 
baryons 
\begin{align}
M_{B,\overline{\bm{3}}}^Q &= M_{\overline{\bm{3}}}^Q + M_{B,\overline{\bm{3}}}^{(1)} +
  M_{B,\overline{\bm{3}}}^{(2)} ,\cr
M_{B,\bm{6}_{1/2}}^Q &= M_{\bm{6}}^Q + M_{B,\bm{6}}^{(1)} +
  M_{B,\bm{3}}^{(2)} -\frac23 \frac{\varkappa}{m_Q},\cr
M_{B,\bm{6}_{3/2}}^Q &= M_{\bm{6}}^Q + M_{B,\bm{6}}^{(1)} +
  M_{B,\bm{6}}^{(2)} + \frac13  \frac{\varkappa}{m_Q}.
\label{eq:FinalMassF}
\end{align}

\begin{table}[htp]
\caption{Results of the masses of the charmed baryon masses in unit of
  MeV. In the third and fourth columns those with the first-order and
  second-order $m_{\mathrm{s}}$ corrections are listed,  The last column
  represents the experimental data.}
\label{tab:2}
\centering 
\begin{tabular}{c c  c  c c}
\hline
\hline
& &  \multicolumn{2}{c} {$m_{\mathrm{s}}$ corrections}  &   \\
${\cal{R}}^{Q}_{J}$& $B_{c}$   & 1st order & 2nd order & Experiment    \\ 
\hline
$\overline{\bm{3}}^{c}_{1/2}$&$\Lambda_{c}$    
& 2274.4 & 2280.7 & 2286.5$\pm$0.1\\ 
$\overline{\bm{3}}^{c}_{1/2}$&$\Xi_{c}$        
& 2481.5 & 2475.2 & 2469.4$\pm$0.3 \\
$\bm{6}^{c}_{1/2}$           &$\Sigma_{c}$     
& 2455.7 & 2448.5 & 2453.5$\pm$0.1 \\
$\bm{6}^{c}_{1/2}$           &$\Xi'_{c}$       
& 2575.2 & 2576.8 & 2576.8$\pm$2.1 \\
$\bm{6}^{c}_{1/2}$           &$\Omega_{c}$     
& 2694.6 & 2700.1 & 2695.2$\pm$1.7 \\
$\bm{6}^{c}_{3/2}$           &$\Sigma^{*}_{c}$ 
& 2523.9 & 2516.7 & 2518.1$\pm$0.8 \\
$\bm{6}^{c}_{3/2}$           &$\Xi^{*}_{c}$    
& 2643.3 & 2645.0 & 2645.9$\pm$0.4 \\
$\bm{6}^{c}_{3/2}$           &$\Omega^{*}_{c}$ 
& 2762.7 & 2768.3 & 2765.9$\pm$2.0 \\
\hline
\hline
\end{tabular}
\end{table}
The numerical results of the charmed baryons are listed in
Table~\ref{tab:2}. As expected, the inclusion of the second-order
$m_s$ corrections produces the results in better agreement with the
experimental data.  It is of interest to compare the present results
with those of Ref.~\cite{Yang:2016qdz}, where the
``\emph{model-independent}'' approach was employed. Theoretically, the
present approach has a certain advantage over
Ref.~\cite{Yang:2016qdz}, since we can consistently 
treat both the valence-quark and sea-quark contributions with the
correct $N_c-1$ factor taken into account. In the model-independent
analysis, an additional scale factor had to be introduced, since it
was not possible to decompose each contribution into the valence and
sea parts~\cite{Kim:2017jpx, Kim:2017khv, Yang:2018uoj}. 

\begin{table}[htp]
\caption{Results of the masses of the bottom baryon masses in unit of
  MeV. In the third and fourth columns those with the first-order and
  second-order $m_{\mathrm{s}}$ corrections are listed,  The last column
  represents the experimental data.} 
\label{tab:3}
\centering 
\begin{tabular}{c c   c  c c}
\hline
\hline
& &  \multicolumn{2}{c} {$m_{\mathrm{s}}$ corrections}  &   \\
${\cal{R}}^{Q}_{J}$ & $B_{b}$  & 1st order & 2nd order & Experiment     \\ 
\hline
$\overline{\bm{3}}^{b}_{1/2}$&$\Lambda_{b}$    
& 5602.7 & 5609.0 & 5619.5$\pm$0.2 \\ 
$\overline{\bm{3}}^{b}_{1/2}$&$\Xi_{b}$        
& 5809.9 & 5803.6 & 5793.1$\pm$0.7 \\ 
$\bm{6}^{b}_{1/2}$           &$\Sigma_{b}$     
& 5812.7 & 5805.5 & 5813.4$\pm$1.3 \\ 
$\bm{6}^{b}_{1/2}$           &$\Xi'_{b}$       
& 5932.1 & 5933.8 & 5935.0$\pm$0.05 \\ 
$\bm{6}^{b}_{1/2}$           &$\Omega_{b}$     
& 6051.6 & 6057.1 & 6048.0$\pm$1.9 \\ 
$\bm{6}^{b}_{3/2}$           &$\Sigma^{*}_{b}$ 
& 5834.7 & 5830.3 & 5833.6$\pm$1.3 \\ 
$\bm{6}^{b}_{3/2}$           &$\Xi^{*}_{b}$    
& 5954.2 & 5958.6 & 5955.3$\pm$0.1 \\ 
$\bm{6}^{b}_{3/2}$           &$\Omega^{*}_{b}$ 
& 6073.6 & 6081.9 &       - \\  
\hline
\hline
\end{tabular}
\end{table}
Table~\ref{tab:3} presents the results of the bottom baryon
masses. Similarly, the second-order $m_s$ corrections improve the
results. The mass of the $\Omega_b^*$ is predicted to be 6081.9 MeV,
whereas the \emph{model-independent} approach of
Ref.~\cite{Yang:2016qdz} predicts $M_{\Omega_b^*}=(6095\pm 4.4)$
MeV. The difference is found to be less than $1\,\%$.  

%==================================
\section{Summay and conclusions}
%==================================
In the present work we investigated the mass spectra of the
lowest-lying singly heavy baryons within the framework of the
self-consistent SU(3) chiral quark-soliton model.  In the model, the
$N_c-1$ light valence quarks polarize the Dirac sea. We obtained the 
soliton energy consisting of the $N_c-1$ valence-quark and sea-quark
energies. Minimizing the soliton energy around the saddle point of the
classical pion field self-consistently, we derived the soliton
mass. Because of the hedgehog symmetry, we embedded the SU(2) soliton
into the flavor SU(3). While we ignore the $1/N_c$ quantum
fluctuations in this mean-field approximation, the rotational zero
modes or rotational $1/N_c$ corrections are taken into account, a
rigid rotation of the soliton being assumed. All the moments  of
inertia were computed in the present work explicitly. 

We consider the effects of flavor SU(3) symmetry breaking to the
second-order in perturbation. As expected, the inclusion of the
second-order $m_{\mathrm{s}}$ corrections leads to the better results of the mass
splittings of both the charmed and bottom heavy baryons than those
with the linear $m_{\mathrm{s}}$ corrections, in comparison with the
experimental data. Having fixed the center mass in each
representation, we were able to obtain the numerical values of all the
lowest-lying singly heavy baryons both in the charm and bottom
sectors. With the second-order $m_{\mathrm{s}}$ corrections included,
the present results are in very good agreement with the experimental
data. The mass of the $\Omega_b^*$ baryon is predicted to be $6081.9$
MeV in the present work. 

%-------------------------------------------------
\section*{Acknowledgments} 
%-------------------------------------------------
We are grateful to M.V. Polyakov and M. Prasza{\l}owicz for valuable
discussion. H.-Ch.K wants to express his gratitude to A. Hosaka for
information on the original paper by W. Pauli and S.M. Dancoff about
the hedgehog Ansatz. H.-Ch. K is also thankful to the members of the
Research Center for Nuclear Physics, Osaka University. 
The work of H.-Ch.K. was supported by Basic   
Science Research Program through the National Research Foundation of
Korea funded by the Ministry of Education, Science and Technology
(Grant Number: NRF-2015R1D1A1A01060707). 

%-------------------------------------------------
\begin{appendix}
\section{Moments of inertia\label{app:A}}
In this Appendix, we compile all relevant formulae for the modified 
$\pi N$ sigma term, the moments of inertia $I_{1,2}$, $K_{1,2}$,
and $N_{1,2}$. All terms consist of the vacuum and sea parts. 
The modified $\pi N$ sigma term is written as 
\begin{align}
\overline{\Sigma}_{\pi N}= \overline{\Sigma}^{\mathrm{val}}_{\pi
  N}+\Sigma^{\mathrm{sea}}_{\pi N},
\label{eq:app1}
\end{align}
where the valence and sea parts are expressed respectively as 
\begin{align}
\bar{\Sigma}^{\mathrm{val}}_{\pi N}  = m_0 (N_{c}-1) \langle
  \mathrm{val} | 
\gamma_{4}  | \mathrm{val} \rangle,\;\;\;
\Sigma^{\mathrm{sea}}_{\pi N}=\frac{m_0}{2}N_{c}\sum_{n}\langle n |
\gamma_{4}  | n \rangle \mathrm{sign}(E_{n}){{\cal{R}}_{{\Sigma}}} (E_n),
\end{align}
where $\gamma_4$ is the Dirac $\gamma$ matrix in Euclidean space
represented as 
\begin{align} 
\gamma_4 = 
  \begin{pmatrix}
    \bm{1} & 0 \\ 0 & -\bm{1}
  \end{pmatrix}.
\end{align}
The function $\mathcal{R}_\Sigma(E_n)$ denotes a regularization
function written as 
\begin{align}
{\cal{R}}_{\Sigma}(E_n)=\frac{1}{\sqrt{\pi}} \int^{\infty}_{0} 
  \frac{du}{\sqrt{u}} e^{-u} \phi(u/E_n^2),
\end{align}
where $\phi(u)$~\cite{Blotz:1992pw} is a cutoff function defined by 
\begin{align}
\phi(u) = c \theta(u-1/\Lambda_1^2) + (1-c)\theta(u-1/\Lambda_2^2).  
\end{align}
The free parameters $\Lambda_1$, $\Lambda_2$, and $c$ are determined
in the mesonic sector by reproducing the pion decay constant
$f_\pi=93$ MeV and the pion mass $m_\pi=139$ MeV. Their numerical
values are explicitly given as $\Lambda_1=381.15$ MeV,
$\Lambda_2=1428.00$ MeV, and $c=0.7276$.  

The moment of inertia tensor $I_{ab}$ is given as 
\begin{align}
I_{\mathrm{ab}} =
  I^{\mathrm{val}}_{\mathrm{ab}}+I^{\mathrm{sea}}_{\mathrm{ab}}, 
\label{eq:momIi}
\end{align}
where 
\begin{align}
I^{\mathrm{val}}_{\mathrm{ab}} &=
  \frac{(N_{c}-1)}{2}\sum_{\mathrm{val,n \ne val}}\frac{\langle n |
  \lambda_{a} | \mathrm{val} \rangle \langle \mathrm{val} |
  \lambda_{b} | n \rangle}{E_{n}-E_{\mathrm{val}}},\cr 
I^{\mathrm{sea}}_{\mathrm{ab}}& = \frac{N_{c}}{4}\sum_{m,n}\langle
  n | \lambda_{a} | m \rangle \langle m | \lambda_{b} | n \rangle
  {\cal{R}}_{I}(E_{n},E_{m}), 
\label{eq:mom_exp1}
\end{align}
with the different regularization function $R_{I}(E_n,\,E_m)$ 
\begin{align}
{\cal{R}}_{I}(E_{n},E_{m}) = \frac{1}{2\sqrt{\pi}}
  \int_{0}^{\infty} \frac{du}{\sqrt{u}} \phi(u)    
\left [ \frac{e^{-u E^{2}_{n}}-e^{-u E^{2}_{m}}}{u(E^{2}_{m}-E^{2}_{n})} 
- \frac{E_{n}e^{-u E^{2}_{n}}+E_{m}e^{-u E^{2}_{m}}}{E_{m}+E_{n}}
  \right ]. 
\end{align}
$\lambda_a$ in Eq.~(\ref{eq:mom_exp1}) denote the Gell-Mann matrices
for flavor SU(3) group, satisfying $\mathrm{tr}(\lambda_a\lambda_b) =
2\delta_{ab}$ and $[\lambda_a,\,\lambda_b]=2if_{abc} \lambda_c$,
$a=1,\cdots,8$. The moments of inertia $I_1$ and $I_2$ are defined by  
\begin{align}
I_{ab} \equiv \left \lbrace \begin{array}{c l}
  I_{1}\delta_{ab} & a,b=1,2,3 \\
  I_{2}\delta_{ab} & a,b=4,5,6,7\\
  0               & a,b=8
\end{array} \right. .
\end{align}
Similarly, the anomalous moments of inertia tensor is expressed as 
\begin{align}
K_{\mathrm{ab}}=K^{\mathrm{val}}_{\mathrm{ab}}+K^{\mathrm{sea}}_{\mathrm{ab}}, 
\end{align}
where
\begin{align}
K_{\mathrm{ab}}^{\mathrm{val}}&=\frac{(N_{c}-1)}{2} \sum_{\mathrm{val,n
  \ne val}}\frac{\langle n | \lambda_{a} | \mathrm{val} \rangle
  \langle \mathrm{val} | \lambda_{b} \gamma_{4} | n
  \rangle}{E_{n}-E_{\mathrm{val}}}, \cr
K_{\mathrm{ab}}^{\mathrm{sea}}&=\frac{N_{c}}{8}\sum_{m,n}\langle
  n | \lambda_{a} | m \rangle \langle m | \gamma_{4}\lambda_{b} | n
  \rangle \frac{\mathrm{sign}(E_{n})-\mathrm{sign}(E_{m}) }
  {E_{n}-E_{m}}.  
\end{align}
The anomalous moments of inertia $K_1$ and $K_2$ are defined by 
\begin{align}
K_{ab} \equiv \left \{ \begin{array}{c l}
  K_{1}\delta_{ab} & a,b=1,2,3 \\
  K_{2}\delta_{ab} & a,b=4,5,6,7\\
  0                & a,b=8
\end{array}   \right. .
\end{align}

Finally we express the third moments of inertia tensor, which
appears only when the second-order $m_{\mathrm{s}}$ corrections are considered. 
\begin{align}
N_{\mathrm{ab}}=N^{\mathrm{val}}_{\mathrm{ab}}+N^{\mathrm{sea}}_{\mathrm{ab}} ,
\end{align} 
then the moment of inertia 
\begin{align}
N^{\mathrm{val}}_{\mathrm{ab}}&=\frac{(N_{c}-1)}{2} \sum_{\mathrm{val,n
  \ne val}}\frac{\langle n | \lambda_{a} \gamma_{4} | \mathrm{val}
  \rangle \langle \mathrm{val} | \lambda_{b} \gamma_{4} | n
  \rangle}{E_{n}-E_{\mathrm{val}}}, \cr
N^{\mathrm{sea}}_{\mathrm{ab}} &= \frac{N_{c}}{4}
  \sum_{m,n}\langle n | \lambda_{a} \gamma_{4} | m \rangle
  \langle m | \lambda_{b} \gamma_{4} | n \rangle
  {\cal{R}}_{N}(E_{n},E_{m}) ,
\end{align}
with the regularization function
\begin{align}
{\cal{R}}_{N}(E_{n},E_{m})=\frac{1}{2\sqrt{\pi}}
  \int_0^{\infty}\frac{du}{\sqrt{u}} 
  \phi(u) \frac{E_{n}e^{-uE^{2}_{n}}-E_{m}e^{-uE^{2}_{m}} }
  {E_{n}-E_{m}}.  
\end{align}
$N_0$, $N_1$, and $N_2$ are defined by 
\begin{align}
N_{ab} \equiv \left \lbrace \begin{array}{c l}
  N_{1}\delta_{ab} & a,b=1,2,3 \\
  N_{2}\delta_{ab} & a,b=4,5,6,7\\
  \frac{1}{3}N_{0} & a,b=8
\end{array}   \right . .
\end{align}

\section{Matrix elements of the SU(3) Wigner $D$
  functions\label{app:b}} 
In Appendix~\ref{app:b}, we tabulate all relevant matrix elements of the
SU(3) Wigner $D$ functions in each representation. 
\begin{table}[htp]
\caption{Matrix elements of the SU(3) Wigner $D$ functions
  $D_{88}^{(8)}$ and $D_{8i}^{(8)} J_i$.}
\begin{tabular}{c c c c c c }
\hline
\hline 
&$ {\cal{R}}$ &T& Y & $ \langle {\cal{R}} Y T J|D^{(8)}_{88}|
{\cal{R}} Y T J \rangle$ &$\langle {\cal{R}} Y T J|D^{(8)}_{8i}J_{i}|
{\cal{R}} Y T J \rangle$   \\ 
 \hline
$\Lambda_{c}$&\multirow{ 2}{*}{$\overline{\bf{3}}$} & 0 & $2/3$ &   $1/4$ & 0 \\ 
$\Xi_{c}$& &$1/2$ &$ -1/3$   &  $-1/8$ & 0 \\ 
\hline
$\Sigma_{c}$&\multirow{ 3}{*}{${\bf{6}}$} &1 & $2/3$   & ${1}/{10}$& $-{\sqrt{3}}/{5}$\\ 
$\Xi_c$        & &${1}/{2}$ &$ -{1}/{3}$& $-{1}/{20}$ & ${\sqrt{3}}/{10}$\\ 
$\Omega_c$  & &0 & $-{4}/{3}$& $-{1}/{5}$ & ${2\sqrt{3}}/{5}$\\  
\hline
\hline
\end{tabular}

\label{tab:4}
\end{table}

\begin{table}[htp]
\caption{Matrix elements of the products of the SU(3) Wigner $D$
  functions, $D_{88}^{(8)} D_{88}^{(8)}$, $D_{8i}^{(8)}
  D_{8i}^{(8)}$ and $D_{8p}^{(8)}D_{8p}^{(8)}$, where index $i$ runs
  from 1 to 3 whereas $p$ does from 4 to 7. }
 
\begin{tabular}{c c c c c c c}
\hline
\hline 
&$ {\cal{R}}$ &T& Y & $ \langle {\cal{R}} Y T J|D^{(8)}_{88}D^{(8)}_{88}|
{\cal{R}} Y T J \rangle$& $\langle {\cal{R}} Y T J|D^{(8)}_{8i}D^{(8)}_{8i}|
{\cal{R}} Y T J \rangle$ & $\langle {\cal{R}} Y T J|D^{(8)}_{8p}D^{(8)}_{8p}|
{\cal{R}} Y T J \rangle$ \\ 
 \hline
$\Lambda_{c}$&\multirow{ 2}{*}{$\overline{\bf{3}}$} & 0 & ${2}/{3}$ & ${7}/{40}$ & ${9}/{40}$ &   ${3}/{5}$\\ 
$\Xi_{c}$& &${1}/{2}$ &$ -{1}/{3}$  & ${1}/{10}$ & ${9}/{20}$ &  ${9}/{20}$\\ 
\hline
$\Sigma_{c}$&\multirow{ 3}{*}{${\bf{6}}$} &1 & ${2}/{3}$  & ${3}/{20}$  & ${19}/{60}$  & ${8}/{15}$\\ 
$\Xi_c$        & &${1}/{2}$ &$ -{1}/{3}$ & ${1}/{10}$  & ${2}/{5}$  & ${1}/{2}$\\ 
$\Omega_c$  & &0 & $-{4}/{3}$& ${1}/{10}$  & ${1}/{2}$  & ${2}/{5}$\\  
\hline
\hline
\end{tabular}
\label{tab:5}
\end{table}

\begin{table}[htp]
\caption{Transition matrix elements of the SU(3) Wigner $D$
  functions  $D_{88}^{(8)}$ and $D_{8i}^{(8)} J_i$, which appear from
  the second-order perturbation. }
\begin{tabular}{c c c c c c c }
\hline
\hline 
&$ {\cal{R}}'$ &$ {\cal{R}}$ &T& Y & $ \langle {\cal{R}}' Y T J|D^{(8)}_{88}|
{\cal{R}} Y T J \rangle$ & $ \langle {\cal{R}}' Y T J|D^{(8)}_{8i}J_{i}|
{\cal{R}} Y T J \rangle$ \\ 
 \hline
$\Lambda_{c}$&\multirow{ 2}{*}{$\overline{\bf{15}}$} 
&\multirow{ 2}{*}{$\overline{\bf{3}}$} & 0 & ${2}/{3}$ & ${3\sqrt{5}}/{20}$ &   0\\ 
$\Xi_{c}$ & & &${1}/{2}$ &$ -{1}/{3}$   & ${3\sqrt{15}}/{40}$ & 0\\ 
\hline
$\Sigma_{c}$&\multirow{ 3}{*}{$\overline{\bf{15}}$} 
&\multirow{ 3}{*}{${\bf{6}}$} &1 & ${2}/{3}$   &${\sqrt{10}}/{10}$ & ${\sqrt{30}}/{15}$\\ 
$\Xi_c$      &  & &${1}/{2}$ &$ -{1}/{3}$& ${\sqrt{15}}/{20}$ & ${\sqrt{5}}/{10}$\\ 
$\Omega_c$&  & &0 & $-{4}/{3}$& 0 & 0\\  
\hline
$\Sigma_{c}$&\multirow{ 3}{*}{$\overline{\bf{24}}$} 
&\multirow{ 3}{*}{${\bf{6}}$} &1 & ${2}/{3}$ & ${1}/{5}$ & $-{\sqrt{3}}/{15}$\\ 
$\Xi_c$      &  & &${1}/{2}$ &$ -{1}/{3}$ & ${\sqrt{6}}/{10}$ & $-{\sqrt{2}}/{10}$\\ 
$\Omega_c$&  & &0 & $-{4}/{3}$& ${\sqrt{6}}/{10}$ & $-{\sqrt{2}}/{10}$\\ 
\hline
\hline
\end{tabular}
\label{tab:6}
\end{table}

\end{appendix}

%=========================================================

\end{document}